\documentclass[aps,pre,amsmath,twocolumn,superscriptaddress]{revtex4-1}
\usepackage[figuresright]{rotating}
\usepackage{braket}
\usepackage{bm}
\usepackage{amsmath,amssymb}
\usepackage{color}
\usepackage{url}
\usepackage{algorithm}
\usepackage{algpseudocode}
\usepackage[titletoc]{appendix}
\usepackage{bbm}
\usepackage{bbold}
\usepackage{mathrsfs}

\usepackage[utf8]{inputenc}

\newcommand{\id}{\mathbb{1}}   

\begin{document}
\title{Unfolding quantum master equation into a system of real-valued equations: \\
computationally effective expansion over the basis of $SU(N)$ generators}

\author{A.~Liniov}
\affiliation{Software Engineering Department, Lobachevsky State University of Nizhny Novgorod, Russia}
\author{I.~Meyerov}
\affiliation{Mathematical Software and Supercomputing Technologies Department, Lobachevsky State University of Nizhny Novgorod, Russia}
\author{E.~Kozinov}
\affiliation{Mathematical Software and Supercomputing Technologies Department, Lobachevsky State University of Nizhny Novgorod, Russia}
\author{V.~Volokitin}
\affiliation{Mathematical Software and Supercomputing Technologies Department, Lobachevsky State University of Nizhny Novgorod, Russia}
\author{I.~ Yusipov}
\affiliation{Department of Applied Mathematics, Lobachevsky State University of Nizhny Novgorod, Russia}
\author{M.~Ivanchenko}
\affiliation{Department of Applied Mathematics, Lobachevsky State University of Nizhny Novgorod, Russia}
\author{S.~Denisov}
\affiliation{Department of Computer Science, Oslo Metropolitan University, N-0130 Oslo, Norway}
\affiliation{Department of Applied Mathematics, Lobachevsky State University of Nizhny Novgorod, Russia}

\begin{abstract}
Dynamics of  an open $N$-state quantum system is typically  modeled with a Markovian master equation describing the evolution of the system's density operator.
By using  generators of $SU(N)$ group as a basis, the density operator can be transformed into a real-valued 'Bloch vector'.
The Lindbladian, a super-operator which serves a generator of the evolution,  
can be expanded over the same basis and recast in the form of a real matrix. 
Together, these expansions result is a non-homogeneous system of $N^2-1$  real-valued linear differential equations for the Bloch vector. 
Now one can, e.g., implement a high-performance parallel  simplex algorithm to find a solution of this system  which  guarantees exact preservation 
of the norm and  Hermiticity of the  density matrix.
However, when performed in a straightforward way, the expansion turns to be an operation of the time complexity $\mathcal{O}(N^{10})$.
The complexity can be reduced  when the number of dissipative operators is independent of $N$, which is often the case for physically meaningful models.
Here we present an algorithm to transform quantum master equation into a system of real-valued differential equations and propagate it forward in time. 
By using a scalable model, we evaluate computational efficiency of the algorithm and demonstrate that it is possible to
handle  the model system with $N = 10^3$ states on a single node of a computer cluster.

\end{abstract}

\maketitle
	
	\maketitle

\section{Introduction}\label{sec:1} The most conventional approach to modeling of the dynamics of an open quantum system, i.e., a system interacting with its environment, is 
to use a Markovian master equation \cite{book,carm}. Such equation describes the evolution of the system's density operator $\varrho$, $\dot{\varrho} = \mathcal{L}(t)\varrho$,
and its key ingredient is the generator of evolution, $\mathcal{L}(t)$, a time-dependent (in general) super-operator \cite{book}.

In order to generate a semi-group and fulfill the condition of complete positivity, this super-operator has to be 
of the so-called Gorini–Kossakowski–Sudarshan–Lindblad (GKSL) form \cite{gorini,lin,chur} (henceforth called `\textit{Lindbladian}'),
\begin{eqnarray}
\nonumber
\mathcal{L}(t)\varrho = \mathcal{L}_H(t)\varrho +  \mathcal{L}_D(t)\varrho  = -i[H(t),\varrho] + \sum_{p=1}^P \gamma_p\mathcal{D}_p(\varrho), \\
\mathcal{D}_p(\varrho) = \frac{1}{2} \Big([L_p,\rho L_p^\dag ]+[L_p \rho,L_p^\dag]\Big),~~~~~~~~~~~~~ 
\label{lind}
\end{eqnarray}
where $H(t)$ is a time-dependent Hamiltonian and the set of quantum
dissipative operators, $\{L_p\}$, 
$p=1,...,P$, capture the action of the environment on the system (formally, 
any complex $N \times N$ matrix could be chosen as an operator $L_p$). 
Dissipative operators act via $P$ `channels' with  non-negative rates $\gamma_p$.
As a theoretical tool, the GSKL equation (1) is very popular in quantum optics \cite{carm}, cavity optomechanics \cite{mar} 
and quantum electrodynamics \cite{QED1,QED2}; it is also used in the context of ultra-cold atom physics \cite{cold1,cold2}. 

When Lindbladian $\mathcal{L}$ is time-independent, the structure of the GSKL equation ensures the existence of an asymptotic state $\varrho^{A}$, which is
a non-trivial zero eigen-element (kernel) of $\mathcal{L}$ \cite{several}.  When the  Lindbladian is time periodic, $\mathcal{L})(t+T) = \mathcal{L}(t)$,
Floquet theory \cite{yakub} applies and the asymptotic density operator is time-periodic
with the same period, $\varrho^{A}(t+T) = \varrho^{A}(t)$  \cite{den1}. In either case, the main challenge consists in explicit numerical evaluation of the matrix form of
operator $\varrho^{A}$ \cite{note}.

Leaving aside recently developed tensor methods \cite{mpo}, which apply to lattice systems \cite{lattice} only, 
there are three  means to find $\varrho^{A}$ numerically. 
Here we only briefly list them (we refer the interested readers to the introduction of Ref.~\cite{qt} for a more detailed discussion).
First, one may use spectral methods (complete/partial diagonalization and different kinds
of iterative algorithms \cite{nation1}) to calculate $\varrho^{A}$ as an eigen-element of $\mathcal{L}$. Next, one can  propagate the system density operator forward in time,
by numerically integrating the GKSL equation, until the operator lands on $\varrho^{A}$. Finally, one can unravel the GKSL equation (\ref{lind}) into a set
of stochastic realizations, called ``quantum trajectories'' (QTs) \cite{zoller,dali,plenio}, and thus transform the
problem into a task of statistical sampling over QTs -- which have to be propagated for a long time in order to approach $\varrho^{A}$ \cite{qt}.

Here we address  the second option; namely, we consider propagation of the density operator forward in time by numerically integrating the GKSL equation.
This strategy was already implemented in a number of  works;  it is also  included in such popular open-source package as QuTiP \cite{qutip}. 
The first step in the realization of this idea  is a vectorization of the density operator,
based either on the straightforward row(column)-wise unfolding of the density matrix or usage of an over-complete basis of matrix units, 
$G_{\beta} \equiv G_{k,l} = |k\rangle\langle l|$,
where $\{|s\rangle\}_{s=1,...,N}$ is a set of basis vectors and a bijection $\beta \Leftrightarrow (k,l)$ is implemented. 
The vectorization renders  the GKSL equation in a system of complex-valued linear differential equitations which is then propagated by using some standard high-order integrators \cite{integrator}.
Neither of the discussed vectorization accounts for the norm conservation, Hermiticty and non-negativity of $\varrho(t)$. At the same time, 
it well known that the first two conditions can be accounted
explicitly \cite{luch} by using an orthonormal (Hilbert-Schmidt)  basis of traceless Hermitian operators, which transform the GKSL equation into a set of \textit{real}-valued linear differential equations \cite{gorini,alicki0,alicki,kimura}. 
For a single qubit this procedure is well-known  as the Bloch-vector representation and it leads to the famous Bloch equations \cite{book}. 
For an $N=3$ system it can be realized by using eight Gell-Mann matrices \cite{gel}.  For any $N > 3$ it can be performed  \cite{alicki0,alicki} by using a complete set 
of infinitesimal generators of the $SU(N)$ group \cite{Li}, rendering density matrix in form of the so-called 'coherence-vector'  \cite{alicki0}. 
However, this strategy was  never implemented in practice 
for  $N > 4$, to the best of our knowledge. We guess that one of the main problems which prevents the usage $SU(N)$ unfolding
is its computational complexity (see Section V). This aspect has not been discussed in the literature; at the same time it is an interesting technical problem, for two reasons,
'physical' and 'computational' ones.


First, the coherence-vector representation allows for an alternative  quantification of entanglement in multipartite systems; see, f.e., Refs.~\cite{ent1,ent2,ent3}.
It also provides a tool to investigate a 'geometry' of quantum states \cite{zuck}  by using the condition of positivity \cite{positivity1,positivity2}.
Second, by performing expansion over the $SU(N)$ generators, a search for $\varrho^{A}$  can be transformed into a standard task of linear programming  \cite{opt}. 
This transformation allows one to use a toolbox of  parallel simplex methods, developed for large optimization problems, and implement them on a cluster or supercomputer \cite{parallel},  
thus opening a way to larger model systems. These reasons are our main motivation. 

In this paper we present an  algorithm which realizes the expansion of a Lindbladian over the basis of $SU(N)$ generators \cite{book1}. 
It is tailored to handle model systems with number $P$ of dissipative channels  which grows sub-linearly with $N$ or remains constant. 
The latter condition is not very limiting; in fact, many currently studied models fulfill it.
One could think, e.g.,  of two 'baths``, $L_1$ and $L_2$, acting at the ends of an 1D spin  chain \cite{znidarich} or a leaking cavity stuffed with a tunable number of qubits which
interact with a photonic mode \cite{cavity}. To illustrate the performance of the algorithm, we use a scalable model, a periodically rocked and dissipatively coupled dimer with
$N-1$ interacting bosons, and demonstrate that the algorithm allows to find expansion for $N = 10^3$ and propagate the obtained system up to time $10T$ 
on a single node in a few hours.

Our work is organized as follows: In Section~\ref{sec:2} we outline
the idea of the expansion and present main definitions.  
In Section~\ref{sec:3} we introduce a scalable model
system.  Section~\ref{sec:5} is devoted to the implementation of the
algorithm on a cluster; its performance and scalability are analyzed in Section~\ref{sec:6}.
These results are summarized, together with an outline of further
perspectives, in Section~\ref{summary}.

\section{Expansion of a Lindbladian over the basis of $SU(N)$ generators} \label{sec:2}

In Refs.~\cite{alicki0,alicki} the expansion is presented in detail; here we only summarize the results and introduce necessary definitions and notations.

The basis consists of $M = N^2-1$ $N \times N$ traceless Hermitian matrices, among which there are \cite{alternatives}

\begin{itemize}
		\item $N(N-1)/2$ symmetric, $S^{(j,k)} = \frac{1}{\sqrt{2}} \left(G_{j,k}+G_{k,j} \right)$, $1 \leq j < k \leq N$,
		\item $N(N-1)/2$ antisymmetric,  $J^{(j,k)} = -\frac{i}{\sqrt{2}} (G_{j,k}-G_{k,j})$, $1 \leq j < k \leq N$, 
		\item and $N-1$ diagonal, $D^l = \frac{i}{\sqrt{l(l+1)}} \bigg( \sum_{k=1}^l {G_{k,k}} - lG_{l+1,l+1}\bigg) $, $1 \leq l \leq N-1$, 
	
\end{itemize}
which are forming a set $\bar{F} = \{F_s\}$, $s=0,...,M$. 
This set is complemented with the identity matrix,  $F_0=\id$.  Now we have a basis which is othonormalized with respect to the trace, $\mathrm{Tr}(F_iF_k)=\delta_{ik}$, and complete. 
Important are commutators and anti-commutators of the basis elements,
\begin{eqnarray}
 [F_i,F_k] = i \sum_{l=1}^{M}{f_{ikl}F_l}, \\
 \{F_i,F_k\} = \frac{2}{N}F_0\delta_{ik} + \sum_{l=1}^{M}{d_{ikl}F_l},
\label{comm}
\end{eqnarray}
with $f_{ikl}$ $(d_{ikl})$ being a real completely antisymmetric (symmetric), with respect to permutation of any pair of indices, tensor,
\begin{eqnarray}
	\begin{matrix}
	f_{mns} = -i \mathrm{Tr}(F_s [F_m,F_n ]), ~~~m, n, s = 1, M \\
	d_{mns} = \mathrm{Tr}(F_s \{F_m,F_n \}), ~~~m, n, s = 1, M
	\end{matrix}
	\label{eq:071}
\end{eqnarray}

A density operator can be expanded over this basis,
\begin{eqnarray}
	\rho = \frac{1}{N}F_0 + \sum_{j=1}^{M}{v_j F_j},
	\label{eq:05}
\end{eqnarray}
where \textit{coherence-vector} (also called 'generalized Bloch vector' or simply 'Bloch vector' \cite{kimura})  $\bar{v}=(v_1,v_2,...,v_M)$ 
consist of real-valued elements \cite{alicki}. As a Hermitian operator, $H(t)$ can also be expanded, 
$H(t) = \sum_{j=1}^{N^2-1}{h_j(t) F_j }$ (without loss of generality, henceforth we assume the Hamiltonian to be traceless). The unitary part of the Lindbladian,  $\mathcal{L}_H$,
yields a $M \times M$ matrix $Q$, with elements
\begin{eqnarray}
q_{sn}=\sum_{m=1}^{M}{f_{mns}h_m}, 
\label{eq:072}
\end{eqnarray}
which is skew-symmetric, $Q^{T} = -Q$, due to the antisymmetry of tensor $f$. Thus,
the unitary part of  the GKSL equation (\ref{lind}) transforms into $\dot{\bar{v}} = Q \bar{v}$.

Expansion of the dissipative part of the Lindbladan is more involved. In the original GKSL equation, this part can be rewritten in the following form \cite{gorini}
\begin{eqnarray}
\mathcal{L}_D = \frac{1}{2}\sum_{j,k=1}^{M} {a_{jk} \left([F_j, \rho F_k^\dag ]+[F_j \rho,F_k^\dag ] \right)},	
	\label{diss_new}
\end{eqnarray}
where complex $M \times M$ matrix $A = \{a_{jk}\}$ is positive semidefinite (at any instant of time), $A \geq 0$, 
and has rank $P$. It can be diagonalized, $\tilde{A} = SAS^\dag = \mathrm{diag}\{\gamma_1, \gamma_2,..., \gamma_P\}$, and  dissipators can be expressed as 
$\bar{L} = S^\dagger\bar{F}$. By using spectral decomposition, $A = \sum_{p=1}^{P} {\bar{l}_p\bar{l}_p^\dag}$, the dissipative part can be recast into
\begin{eqnarray}
\mathcal{L}_D = \frac{1}{2}\sum_{p=1}^{P} \gamma_p  \sum_{j,k=1}^{M} {l_{p;j} l^{\ast}_{p;k}\left([F_j, \rho F_k^\dag ]+[F_j \rho,F_k^\dag ] \right)}.	
	\label{diss_new}
\end{eqnarray}

To the equation for the coherence-vector $\mathcal{L}_D$ contributes with $M \times M$ matrix $R$ and vector $K$,
\begin{eqnarray}
	\frac{dv(t)}{dt} = [Q(t)+R ]v(t)+K,
	\label{eq:06}
\end{eqnarray}
with components 
\begin{eqnarray}
	r_{sm}= ~~~~~~~~~~~~~~~~~~~~\\ 
	\nonumber
	-\frac{1}{2} \sum_{p=1}^{P} \gamma_p \sum_{j,k,l=1}^{N^2-1}{l_{p;j} l^{\ast}_{p;k} (z_{jlm} f_{kls} + \overline{z_{klm}} f_{jls})}, \\
	\nonumber
	m, s = 1,...,M
	\label{eq:074}
	\end{eqnarray}
\begin{eqnarray}
	k_s=\frac{i}{N} \sum_{p=1}^{P} \sum_{j,k=1}^{N^2-1}{l_{p;j} l^{\ast}_{p;k} f_{jks}}, ~~~s=1,...,M
	\label{eq:073}
\end{eqnarray}

Summation over $p$ in Eqs.~(10 - 11)
renders a trivial parallelization, 
so henceforth we restrict consideration to the case $P=1$ (a single dissipative operator).

\section{Testbed model} \label{sec:3}
	
As a testbed we use a model describing $N-1$ indistinguishable interacting bosons, which are hopping between the sites of a periodically modulated dimer. 
The model is described with a time-periodic Hamiltonian 
	\begin{eqnarray}
	\nonumber
	H(t) = J(b_1^\dag b_2+b_1 b_2^\dag ) + ~~~~~~~~~~~~~~~~~~~~~~~~~\\
	~~~~~~~~~~\frac{2U}{(N-1)} \sum_{j=1}^2{n_j (n_j-1)}+ 
	\varepsilon(t)(n_2 - n_1)
	\label{dimer_hamiltonian}
	\end{eqnarray}
where $b_j$ and $b_j^\dag$ are the annihilation and creation operators of an atom at site $j$, 
while $n_j = b_j^\dag b_j$ is the operator of number of particle on $j$-th site, $J$ is the tunneling amplitude, $U/(N-1)$ is the interaction strength (normalized by a number of bosons), and $\varepsilon(t)$ 
represents the modulation of the local potential. $\varepsilon(t)$ is chosen as $\varepsilon(t) = \varepsilon(t + T) = E + A\theta(t)$, 
where $E$ is the stationary energy offset between the  sites and $A$ is the dynamic  offset. This type of Hamiltonian has been studied theoretically \cite{weiss, vardi, trimborn, poletti} 
and was implemented in several experiments \cite{gross, tomkovic}.
Two types of the driving are are popular: (i) piecewise constant periodic driving, $\theta(t) = 1$ for $0 \leq t < T / 2$, $\theta(t) = -1$ for $T / 2 \leq t < T$ and (ii) 
sinusoidal driving, $\theta(t) = \sin(t)$. \par

As a dissipative operator we use 
	\begin{eqnarray}
	L=\frac{\gamma}{N-1} (b_1^\dag + b_2^\dag )(b_1 - b_2).
	\label{eq:3_1}
	\end{eqnarray}
This dissipative coupling tries to `synchronize' the dynamics on the sites by constantly recycling
anti-symmetric out-phase mode into symmetric in-phase one \cite{DiehlZoller2008}. Since the jump operator is non-Hermitian, the asymptotic state is different (in general) from  the maximally mixed state,
$\varrho^{A} \neq \id/N$. As a result of modulations, the asymptotic state is characterized by a time-periodic density operator, $\varrho^{A}(t+T)=\varrho^{A}(t)$, so that
the asymptotic state has to be specified over one period of the driving, $\varrho^{A}(t_s)$, $t_s = t~ \mathrm{mod}~ T \in [0,T]$ \cite{den1}.

\par 

Note that term $H^J=(b_1^\dag b_2+b_2^\dag b_1)$ is represented by a tridiagonal matrix (in the Fock basis).  
The components $H^U=\frac{2U}{N-1} \sum_{j=1}^2{n_j (n_j-1)}$ and $H^E=\varepsilon(t)(n_2-n_1 )$ are diagonal matrices. 
Thus, the Hamiltonian is represented by a symmetric tridiagonal matrix. The dissipator $L(t)$ is an antisymmetric tridiagonal matrix.

To visualize the state of this many-body system we use the idea of quantum  'bifurcation diagram' introduced in Ref.~\cite{ivanchenko}. 
The diagram shows (on $y$-axis) the probability to find -- at time $t_s=0$ -- exactly $n$ bosons on the first site as a function of the interaction strength $U$ ($x$-axis).
The probabilities are obtained as the diagonal elements of the density operator $\varrho^{A}(0)$ expressed in the Fock basis.

\section{Implementation} \label{sec:5}

The expansion described in Section \ref{sec:2}, together with the propagation,  can be implemented in four  steps, see Table \ref{tab:01}. 

During the initialization step, the algorithm  reads initial data, allocates memory, 
and initializes main data structures. During the second step, it prepares data for subsequent calculations. 
Namely, the coefficients of the expansion of the matrices $H$ and $L$ in the basis $\{F_i\}$, 
and the coefficients of the ODE system, Eq.~(\ref{eq:06}), are calculated. In the third step, the ODE 
system is integrated up to time $t$. Finally, during the finalization step, the computed results are saved to files and  memory is released.
	
	\begin{table*}[t]
		\centering
		\caption{Main algorithm}
		\label{tab:01}
		\begin{tabular}{ll}
			\hline
			\textbf{Step}       & \textbf{Substep}                                                                                                                                                                                                                                                                                                                                                                                                           \\ \hline
			1. Initialization   & \begin{tabular}[c]{@{}l@{}}1.1. Read the initial data from configuration files.\\ 1.2. Allocate and initialize memory.\end{tabular}                                                                                                                                                                                                                                                                                    \\ \hline
			2. Data preparation & \begin{tabular}[c]{@{}l@{}}2.1. Compute the coefficients $h_j$, $l_j$ of the expansion of the matrices $H$ and $L$ in the basis $\{F_i\}$.\\ 2.2. Compute the coefficients $f_{mns}$, $d_{mns}$, $z_{mns}$ by formulas (\ref{eq:071}).\\ 2.3. Compute the coefficients $q_{sm}$ by formula (\ref{eq:072}).\\ 2.4. Compute the coefficients $k_s$ by formula (\ref{eq:073}).\\ 2.5. Compute the coefficients $r_{sm}$ by formula (\ref{eq:074}).\\ 2.6. Compute the initial value $v(0)$.\end{tabular} \\ \hline
			3. ODE integration  & \begin{tabular}[c]{@{}l@{}}3.1. Integrate the ODE (\ref{eq:06}),over time to $t = T$ by means of the Runge-Kutta method.\\ 3.2. Compute $\rho(T)$ by formula (\ref{eq:05}).\end{tabular}                                                                                                                                                                                                                                                              \\ \hline
			4. Finalization     & \begin{tabular}[c]{@{}l@{}}4.1. Save the results.\\ 4.2. Release memory.\end{tabular}                                                                                                                                                                                                                                                                                                                                    \\ \hline
		\end{tabular}
	\end{table*}

The implementation of the expansion (Step 2) seems to be straightforward but a brute force direct realization 
leads to a high time complexity and memory requirements, even in the case of  sufficiently sparse Hamiltonians and dissipator matrix. 
Here we propose an implementation that allows to substantially reduce  memory requirements and time complexity. 
This is  achieved by taking into account sparsity patterns of the involved  matrices and performing operations only with nonzero elements. 
In this section we estimates the implementation complexity and 
the amount of memory required for the general case of dense matrices (both of a Hamiltonian and dissipators), 
as well as for the application considered in Section \ref{sec:3}. Note that all operator matrices have size $N$ and we use $NZ$ to denote the number of nonzero elements in them.\par

\vspace{.4cm}

	\textbf{Step 1. Initialization.}\par
	
	The initialization of the Hamiltonian and 
	the dissipator matrices  requires $O(N^2)$ operations and $O(N^2)$ space in general case (for dense matrices). 
	When dealing with the dimer model, we use the sparse matrix storage format CSR, which requires $O(N)$ operations for initialization and $O(N)$ space.\par
	
	\vspace{.4cm}
	
	\textbf{Step 2. Data preparation.}\par
	
	First, we need to compute the coefficients $h_j$, $l_j$ of 
	the expansion of the matrices $H$ and $L$ in the basis $\{F_i\}$ (step 2.1.). 
	The coefficients of the elements of the basis $S^{(j,k)}$, $J^{(j,k)}$ are calculated for $O(1)$ operations, thanks to the form of the basis matrices, 
	which gives the total time complexity $O(NZ_H+NZ_L)$. 
	The coefficients for all $D^lD$ are calculated with $O(N)$ operations. 
	Thus, the total time complexity is $O(NZ_h+NZ_l+N)$. We store the coefficients in two arrays. 
	The first array contains the values and takes $O(NZ)$ space. 
	The second array represents the expansion and contains $-1$ for zero coefficients and indexes in the first array for nonzero ones. 
	It takes $O(N^2)$ space. This allows quickly accessing the element and checking if it is equal to zero. In the case of dense matrices $NZ_h=NZ_l=N^2$, while for the dimer
	model $NZ_h=NZ_l=3N-2$.\par
	
\begin{figure}
		\includegraphics[width=0.47\textwidth]{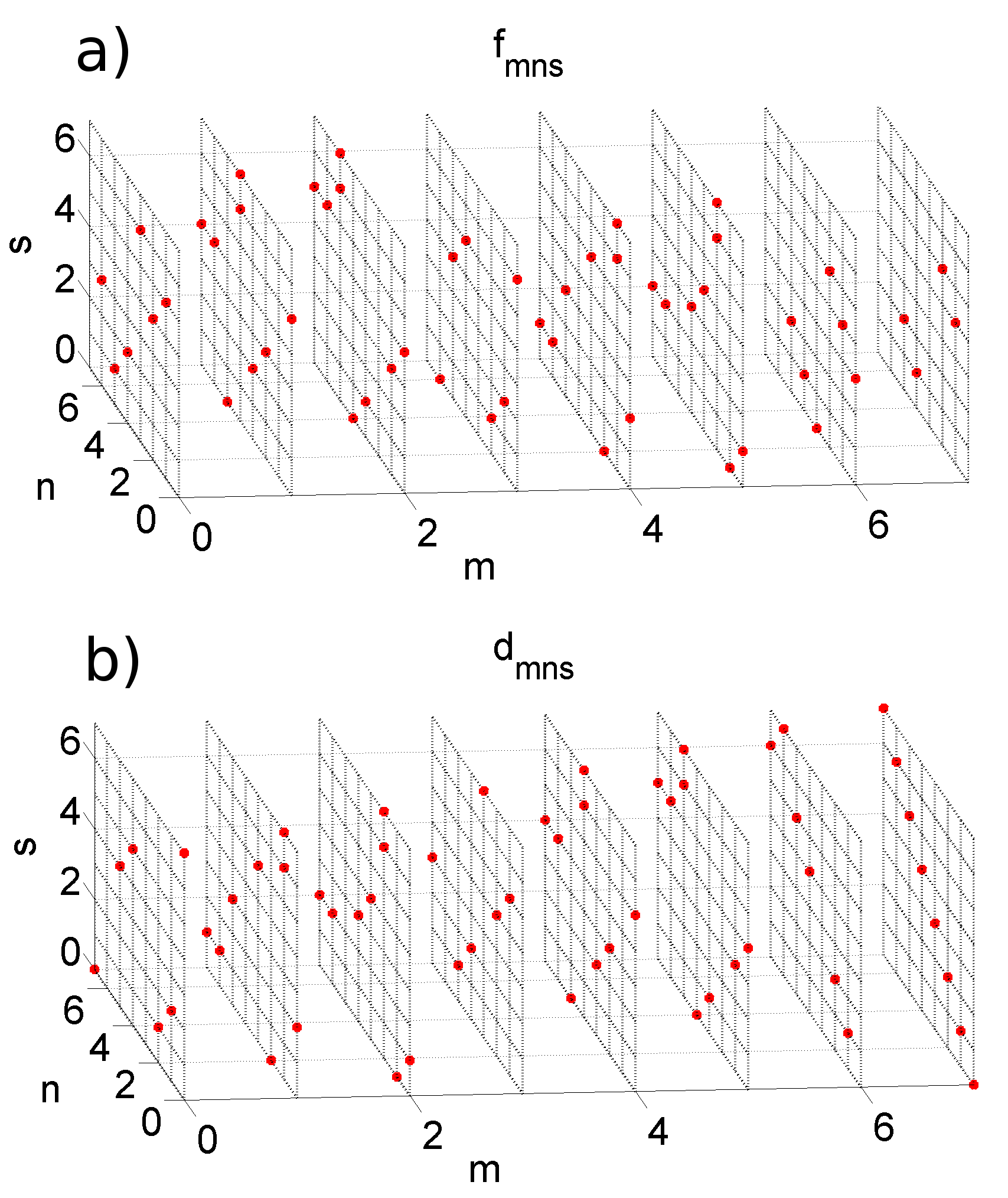}
		\caption{\label{fig:01} Sparsity patterns of tensors $\{d_{mns}\}$ (a) and $\{f_{mns}\}$ (b) for $N = 3$. The red points indicate nonzero elements.}
	\end{figure}

	Next, we need to compute the coefficients $f_{mns}$, $d_{mns}$, $z_{mns}$ (step 2.2.). 
	The number and values of these coefficients depend only on the size of the problem, $N$. Their direct calculation by Eq.~(\ref{eq:071}) requires  
	$2(N^2-1)^3$ matrix multiplications for the basis matrices $\{F_i\}$. 
	Most multiplications require a fixed number of operations independent of $N$. Multiplication with the participation of the matrices $\{D^l\}$ require up to $O(N)$ operations. 
	The total time complexity is therefore $O(N^6)$. 
	It can be significantly reduced by taking into account the sparsity patterns of the matrices $\{F_i\}$. 
	The main idea is to account for nonzero coefficients only. We found that it is possible to determine the set of nonzero coefficients analytically. 
	Namely, the number of nonzero coefficients $f_{mns}$ is $NZ_F=5N^3-9N^2-2N+6$, the number of nonzero coefficients $d_{mns}$ is $NZ_D=6N^3-N(21N+7)/2+1$, 
	and the complexity of calculating each coefficient is $O(1)$. Thus, the overall time complexity is $O(N^3)$. 
	It should be noted that, despite the apparent uniform distribution of $O(N^3)$ nonzero coefficients in tensors of size $N^2 \times N^2 \times N^2$, 
	every set of two-dimensional sections of the tensor $\{d_{mns}, m=\mathrm{const}\}$, $\{d_{mns}, n=\mathrm{const}\}$, $\{d_{mns}, s=\mathrm{const}\}$ includes $O(N)$ two-dimensional sections, with
	$O(N^2)$ elements in each of them. It results in $O(N^3)$ elements in total in every such sub-tensor. 
	The example of nonzero coefficients distribution for $N=3$ is shown in Fig.~\ref{fig:01}).
	
	When calculating the coefficients, we use the coordinate sparse 
	matrix format, which requires $O(NZ_F + NZ_D)$ space. Next, 
	we convert the tensors from the coordinate format to the CRS format. For this we 
	employ the quicksort algorithm to dictionary sort triples of indices $(m,n,s)$. 
	The resulted complexity of this step is $O([NZ_f + NZ_d]\log[NZ_f+NZ_d]) \sim O(N^3 \log N)$ operations and it requires $O(N^3)$ space. 
	The subsequent addition of the tensors $Z = F + D$ has the time complexity $O(NZ_f + NZ_d) \sim O (N^3)$ and requires another $O(N^3)$ space. 
	Two points are of importance.  First, it is possible to pre-compute the tensor $Z$ and store it in a file. 
	Second, we can get rid of memory allocation and element computation for the tensors $D$, $F$, and $Z$. 
	Instead, elements of these tensors can be computed on fly, when needed. We use this approach to decrease memory consumption.\par
	
	The next part of the algorithm (step 2.3.) computes coefficients $q_{sm}$, Eq.~(\ref{eq:072}). 
	A straightforward implementation requires $O(N^6)$ operations. 
	However, we again can significantly decrease the computational load thanks to sparsity of the tensor $F$. Computations can be done by using the following recipe: \par
	
	\begin{enumerate}
		\item Represent the tensor $F$ in the coordinate format $(F')$ in which only nonzero $f_{nms}'=f_{nms}*h_n$ are stored. 
		It takes $O(N^3)$  time and $O(NZ_{F'})$  space. 
		In the general case, $NZ_{F'}$ depends essentially on the form of the Hamiltonian. If there are non-zero elements 
		on its main diagonal, $NZ_{F'} \sim O(N^3)$, otherwise $O(N\cdot NZ_H)$.
		\item Sort $F'$ elements by a pair of indices $(s, m)$. It can be done by using  the counting sort (or radix sort) algorithm, which results in $O(N^3)$ scaling in time 
		and $O(NZ_{F'})$ in space.
		\item Calculate  sums $q_{sm}=\mathrm{Re}(\sum_{n=1}^{N^2-1}{h_n f_{nms} })$ and form the $Q$ matrix in the CRS format. 
		The matrix can be filled in two steps. At the first, the number of nonzero elements in each row is computed. 
		Next, the elements are calculated and indexes are written. All this can be done in $O(NZ_{F'})$ time and requires $O(NZ_{F'})$ space.
		The resulted time and space complexity of step 2.3 is $O(N^3)$. 
		The number of nonzero elements in the resulting matrix $Q$ depends essentially on the form of the Hamiltonian, but it does not exceed $O(N^3)$.
	\end{enumerate}
	
	During  the next step (2.4.), we compute the coefficients $k_s$ by formula (\ref{eq:073}).\par

	\begin{table*}[t]
		\centering
		\caption{Algorithm complexity}
		\label{tab:02}
		\begin{tabular}{llcclcc}
			\hline
			&  & \multicolumn{2}{c}{\textbf{Complexity}}        &  & \multicolumn{2}{c}{\textbf{Complexity}}          \\
			\textbf{Algorithm step} &  & \multicolumn{2}{c}{\textbf{for dense $H$ and $L$}} &  & \multicolumn{2}{c}{\textbf{for the dimer model}} \\
			&  & \textbf{Time}            & \textbf{Space}      &  & \textbf{Time}             & \textbf{Space}       \\ \cline{1-1} \cline{3-4} \cline{6-7} 
			1. INITIALIZATION       &  & $O(N^2)$                 & $O(N^2)$            &  & $O(N)$                    & $O(N^2)$             \\
			2. DATA PREPARATION     &  & $O(N^5  \log N)$        & $O(N^4)$            &  & $O(N^3  \log N)$         & $O(N^3)$             \\
			3. ODEINTEGRATION       &  & $O(N^4)$                 & $O(N^2)$            &  & $O(N^3)$                  & $O(N^2)$             \\
			4. FINALIZATION         &  & $O(N^2)$                 & –                   &  & $O(N^2)$                  & –                    \\ \hline
		\end{tabular}
	\end{table*}

	A direct calculation of the coefficients requires $O(N^4)$ operations. This can be decreased if vector $l$ is sparse. 
	To do this, we convert the vector into the coordinate format,  find all nonzero $f_{jks}$ for each nonzero $l_j \overline{l_k}$,
	and add $l_j \overline{l_k}f_{jks}$ to corresponding $k_s$. It requires $O(NZ_l^2)$ time. 
	The same result can be achieved by using the sparsity of the tensor $F$. 
	Thus, we can just go through all nonzero elements of $F$, adding $l_j \overline{l_k}f_{jks}$ to corresponding $k_s$. 
	It requires $O(N^3)$ time. The choice of the algorithm is determined by the relation between $N$ and $NZ_l$. 
	We use the second option because it is independent of the input data. In any case, storing the vector $k_s$ requires $O(N^2)$ space.\par

	\begin{figure*}
		\includegraphics[width=0.8\textwidth]{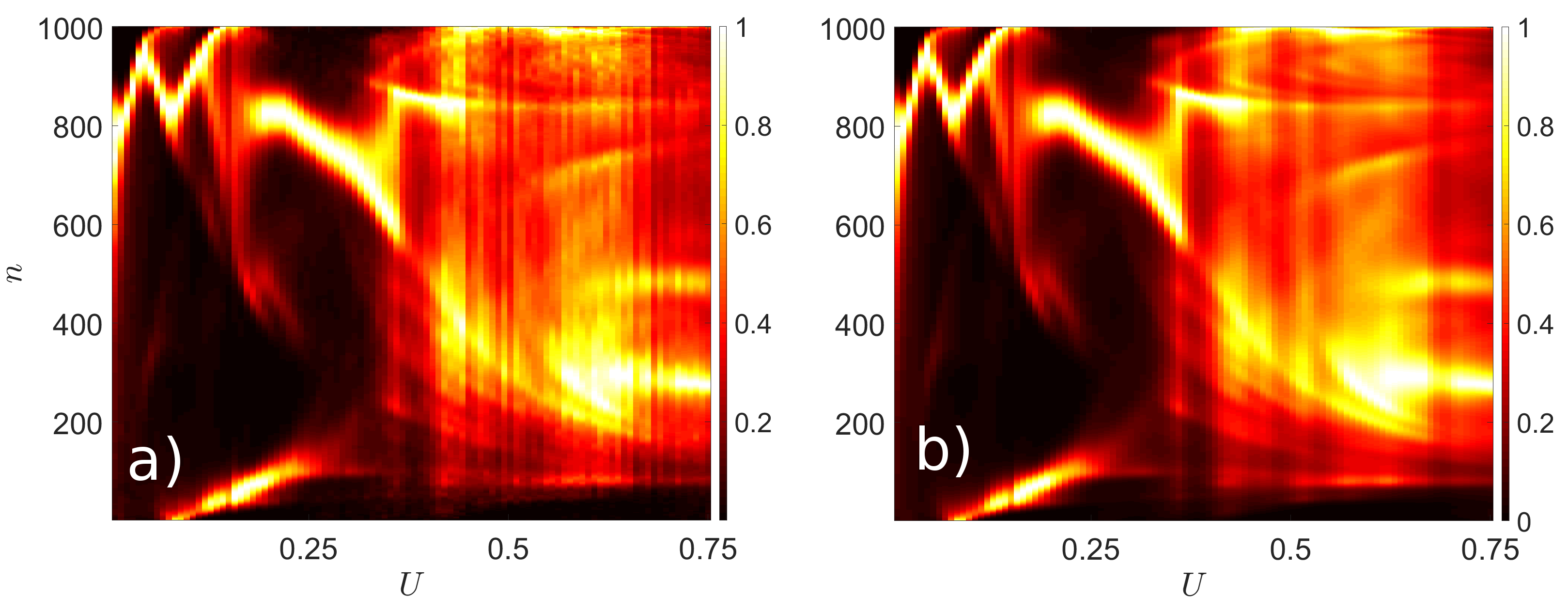}
		\caption{\label{fig:02} Bifurcation diagrams for the modulated dimer with $999$ bosons, Eq. ~(\ref{dimer_hamiltonian}),		
		obtained with (a) the QT implementation~\cite{qt} and (b) with the proposed algorithm.
		To calculate probabilities to find $n$ bosons on the first site ($y$-axis) with the QT method,  $8000$ realizations were sampled for each value of $U$. 
		Parameters of the model are  $J = -1$, $E = 1$, $A = 1.5$, $T = 2 \pi$, and $\gamma = 0.1$. For every value of $U$, the probabilities shown on $y$-axis are divided
		with the maximal probability. The initial state is $\varrho(0) = |0\rangle\langle0|$.}
	\end{figure*}

	Next (step 2.5), we compute the coefficients $r_{sm}$ by using Eq.~(\ref{eq:074}).	
	Again, a straightforward calculation of the coefficients  -- even for $P=1$ -- requires $O (N^{10})$ operations, which is unacceptable. 
	The following details should to  be taken into account in order to reduce the scaling:
	
	\begin{enumerate}
		\item Tensors $F$ and $Z$ are sparse and contain $O(N^3)$ elements each;
		\item Tensors $F$ and $Z$ are filled in such a way that their two-dimensional 'sections' (matrices)  contain $O(N)$ to $O(N^2)$ elements;
		\item Vector $l$ is sparse if the dissipator is sparse. For the dimer model this vector contains $NZ_l = 3N-2$ nonzero elements ($N^2-1$ elements in the general case).
	\end{enumerate}
	
	The corresponding algorithm reads:
	
	\begin{enumerate}
		\item Convert tensors $F$ and $Z$ to the coordinate format ($F'$ and $Z'$ correspondingly). 
		Both tensors store only nonzero elements $\overline{l_i}f_{ijk}$ and $l_i * z_{ijk}$. It can be done in $O(N^3)$ time and space. \par
		
		If matrix $L$ is dense, tensors $F'$ and $Z'$ contain $NZ_{F'}$, $NZ_{Z'} \sim O(N^3)$ nonzero elements, and the two-dimensional sections of $F'$ and $Z'$ contain $O(N^2)$ nonzero elements.\par
		
		Thanks to the sparsity of matrix $L$, the number of nonzero elements is much smaller in the dimer model. Namely, it is $O(N^2)$ for $F'$ and $Z'$ and $O(N)$ for their $2D$ sections.
		
		\item Sort elements of $F'$ and $Z'$ on the second and the third indexes. When using the counting sort or radix sort algorithm, it takes $O(NZ_{F'}+NZ_{Z'})$ time and space.
		
		\item Compute sums of elements with the same second and third indices. The results can be represented as  matrices $F''$, $Z''$ in the coordinate format, ordered by the second and third coordinates. It requires $O(NZ_{F'}+NZ_{Z'})$ time and space.\par 
		
		If the matrix $L$ is dense, $O(N)$ rows in the matrices $F''$ and $Z''$ contain $O(N^2)$ nonzero elements. The remaining $O(N^2)$ rows can contain $O(N)$ nonzero values. In the application considered in this paper all rows of the matrices $F''$ and $Z''$ contain no more than $O(N)$ nonzero elements.
		
		\item Store the matrix $R$ as an array of the red-black trees where every row of the matrix is represented as a separate tree. 
		For each $l= 1,...,M$ compute all products $l_j \overline{l_k}(z_{jlm} f_{kls} + \overline{z_{klm}}f_{jls})$ and add the results to the corresponding elements of the matrix R.\par 
		
		It can be done in $O(N\cdot(N^2)^2 \cdot \log N)+O(N^2 \cdot N^2 \cdot \log N) \sim O(N^5 \log N)$ time and $O(N^4)$ space for a dense matrix and $O(N\cdot(N)^2  \log N)$ 
		time and -- $O(N^3)$ space for the dimer model.
		
		\item Convert matrix $R$ to the CRS format. It requires $O(N^4  \log N)$ time and $O(N^4)$  space in the general case, and 
		$O(N^3  \log N)$ time and $O(N^3)$ space for the model problem.\par
		
		Consequently, the step 2.5 requires $O(N^5  \log N)$ time and $O(N^4)$ space for the general case, and $O(N^3  \log N)$ time and $O(N^3)$ space for the dimer model.
	\end{enumerate}
	
	Finally, during step 2.6. we compute  initial coherence-vector $v(0)$ and then initiate time propagation. 
	For this purpose, we expand the initial state $\rho(0)$ in the $F$-basis. It takes $O(N^2)$ time and $O(N^2)$ space (see explanations for step 2.1.).\par

	\vspace{.4cm}
	
	\textbf{Step 3. ODE integration.}\par
	
	During this step we integrate the linear real-valued ODE system, Eq.~(\ref{eq:06}), over time by $t = T$ (step 3.1.) 
	and compute resulted $\rho(T)$ (step 3.2.). The complexity of the ODE integration is determined by the method used and the number of nonzero elements 
	in  matrices  $Q$ and $R$ (up to $O(N^4)$ elements for dense matrices). For example, the time complexity of one time step is $O (N^4)$ for the Runge-Kutta integration. 
	However, the time complexity of one step is $O(N^3)$ for the dimer model, Section \ref{sec:3}. The integration of the corresponding  ODE system 
	by the forth-order Runge-Kutta method requires $O(N^2)$ additional space for storing intermediate results. The computation of $\varrho(T)$ has complexity $O(N^2)$, both in time and space.\par
	
	\textbf{Step 4. Finalization.}\par
	
	During this step we save results to files and release memory. The time complexity is $O(N^2)$.\par
	
	Resulted time and space complexity estimations are presented in Table \ref{tab:02}.

	\section{Performance analysis}\label{sec:6}
	
	For performance tests we use a node of the Lobachevsky supercomputer \cite{top} with a $2 \times8$-core Intel Xeon CPU E5-2660, 2.20GHz, 
	128 GB RAM. The code was compiled with Intel C++ Compiler, Intel Math Kernel Library and Intel MPI from the Intel Parallel Studio XE suite of development tools.

\begin{figure}
		\includegraphics[width=0.48\textwidth]{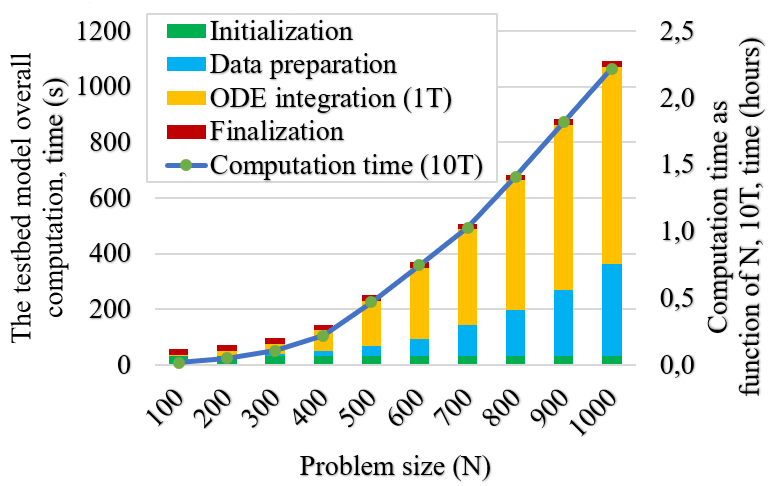}
		\caption{\label{fig:04} Computation time as a function of $N$. The total computation time (line; right y-axis) was measured for the propagation time $10T$. More detail analyis
		was performed for the propagation time $T$ (bars; left y-axis). To get estimates for longer propagation times, one has to scale linearly computation time of the ODE integration step. }
	\end{figure}

	To start, we check (simply visually) the correctness   of the algorithm by comparing its results with the results of  our recently proposed QT implementation \cite{qt}.
        For that we calculate bifurcation diagrams for the dimer model with $N-1=999$ bosons after the transient time $t=10T$, see Fig. \ref{fig:02}. The advantage of the QT implementation 
        is its speed: the scanning over different values of $U$ can be performed faster (even for longer transient time). However, its main disadvantage is the accuracy -- the integration 
        of the ODEs by using the fourth-order Runge–Kutta is much better in this respect [compare Fig.~\ref{fig:02}(a) to Fig.~\ref{fig:02}(b)] and it allows to avoid statistical sampling.
	The computational performance of the algorithm as the function of $N$ is shown  in Fig. \ref{fig:04} (line). 	Note that for $N = 10^3$ it takes less than $3$ hours. 
	
	Next we analyze scaling of computation times of different steps as functions of  $N$. To do so we set propagation time to $T$.
	The results are shown in Fig. \ref{fig:04}(bars). First, it shows that the time of the preparation step, although significant, is substantially smaller than the time of ODE integration.
	Taking into account that the preparation step is performed once, while integration time scales linearly with the actual time of propagation, we conclude that it is the latter 
        that  determines the total computation  time  of the algorithm. It is evident that the initialization and finalization steps
        do not make a significant impact on the overall computational time.

\begin{figure}[b]
		\includegraphics[width=0.48\textwidth]{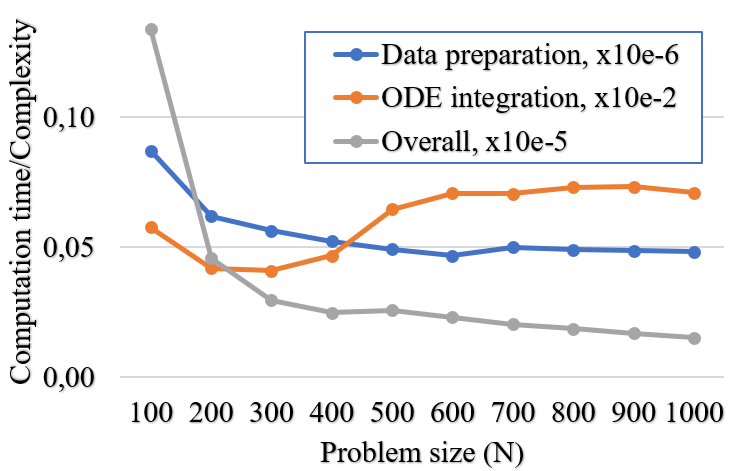}
		\caption{\label{fig:06} Ratio between the actual computation time and asymptotic prediction. 
		For a fixed $N$, the ratio is scaled in order to make the total time (to perform all steps) equal to one.}
	\end{figure}

\begin{figure}
		\includegraphics[width=0.48\textwidth]{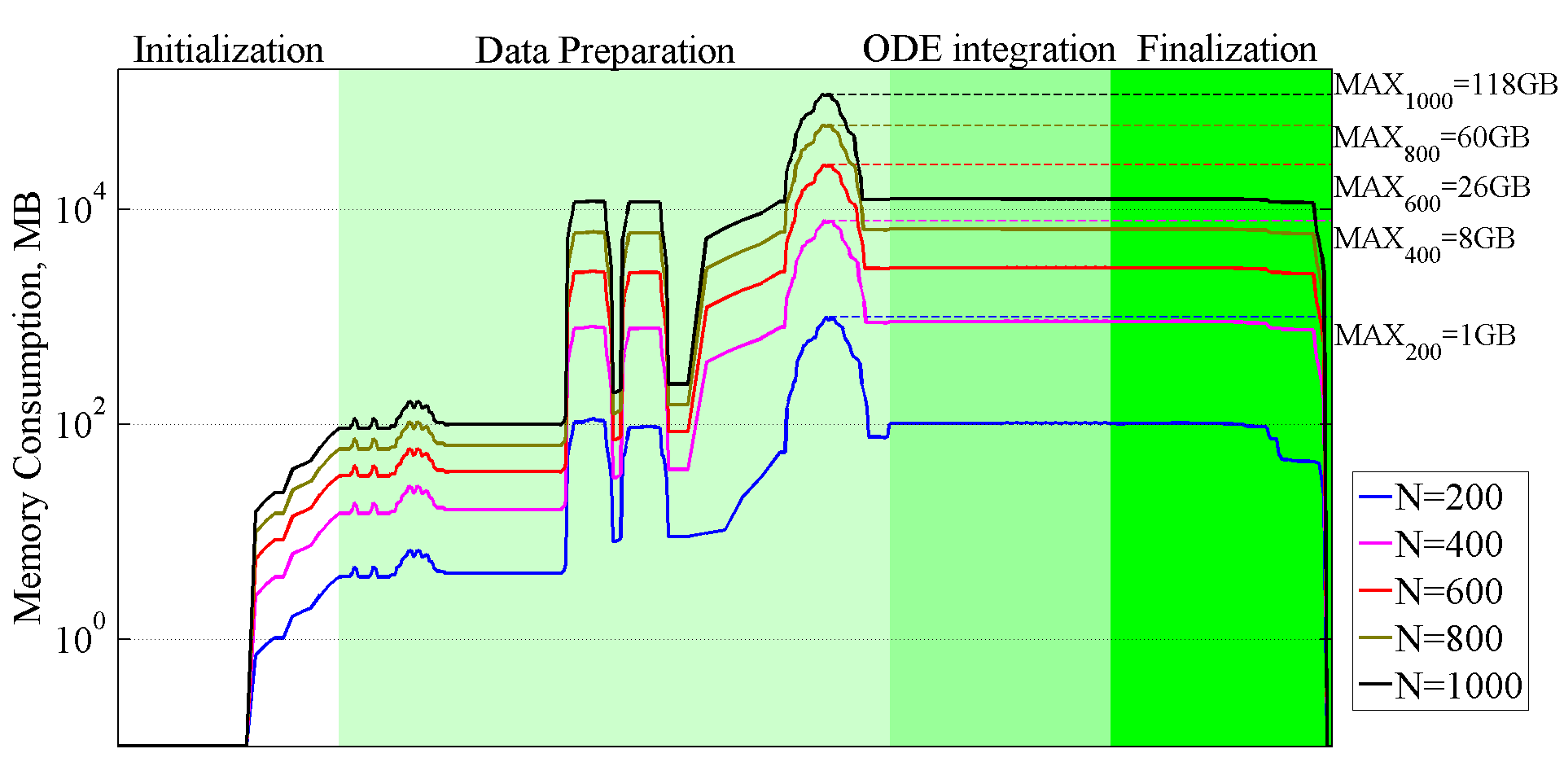}
		\caption{\label{fig:07} Scaling of memory used during the different steps of the algorithm  as functions of  $N$, from $N=200$ (blue line) to $N=10^3$ (black line).}
	\end{figure}

	Now we  compare theoretical predictions obtained for $N=10^2$ for the propagation time $10T$ upon the increase of the problem size to $N=10^3$, see Fig. \ref{fig:06}.
	Namely, we compare computation times of the most time consuming steps and the overall  computation time with the theoretical predictions. To do so, we scale (as discussed above)
	the measured estimate and compare them with actual ones. First observation is that the relation obtained for the preparation step saturates to a constant value. 
	The relation for the integration step 
	slowly goes down with the increase of $N$; therefore, the estimated obtained early can be considered as an upper bound and the actual number 
	of the operations during this step is less then expected. It is not a surprise if we recall that the matrix sparsity scales non-trivially with $N$. 
		Finally, we analyze the scaling of the memory use. First we consider how it scales with the propagation time. The results of the analyses are presented with 
	Fig. \ref{fig:07}, where the memory used during different algorithm steps is shown as function of $N$. It peaks during the data preparation step, when matrix
	$Q$ is calculated. It is noteworthy that the memory use is around  $100GB$ for $N = 10^3$; this already sets certain demands to the computational cluster.\par	
\begin{figure}[b]
		\includegraphics[width=0.48\textwidth]{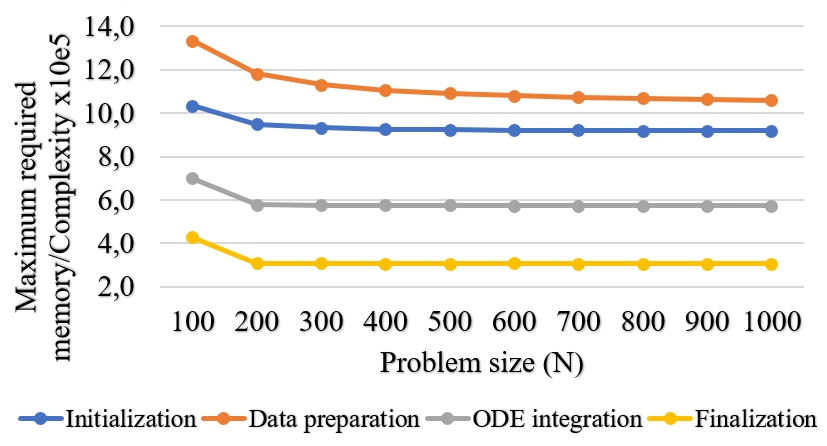}
		\caption{\label{fig:08} Ratio between the maximal memory use measured in computation experiments and the corresponding  theoretical estimates.  
		For a fixed $N$, the ratio is scaled in order to make the total time (to perform  all steps) equal to one.}
	\end{figure}

	Now we compare the results of computation experiments with theoretical estimates. For this we tune the size of the model from $N =10^2$ to $10^3$ 
	and calculate the ratio between the maximal memory use obtained in numerical experiments and estimates, see Fig. \ref{fig:08}). Thus, the latter are confirmed.

	\section{Conclusions}\label{summary}
	
	In this work we presented an algorithm to transform quantum Markovian master equations in the Gorini–Kossakowski–Sudarshan–Lindblad form into systems of $N^2-1$ 
	linear real-valued ordinary differential equations. 
	We included a propagation  step into the algorithm so that a model system can be integrated forward in time and its asymptotic state can be approached. 
	We evaluated the performance of the algorithm and demonstrated that it is possible to simulate a quantum model with $N=10^3$ states and propagate it in time
	 on the medium-size cluster on time scale of several hours.

	Alternatively, the asymptotic state can be calculated by using the method of linear programming \cite{opt}, by 
	minimizing a norm of the rhs of the  ODE system. 
	This open a news volume for parallelization since there is a toolbox of parallel methods designed to solve such minimization problems \cite{parallel}.

	We see the proposed algorithm as a computational mean to explore open quantum models and search for footprints of dissipative 'Quantum Chaos' \cite{QS}. 
	Namely, it could be used to grasp the asymptotic (or a near asymptotic) density operator
	of an open system with several thousands of states so that we can analyze spectral properties of the operator. 
	It could be also used to explore the issue of quantum stability, which for open systems was recently discussed in Refs.~\cite{stab1,stab2}.

A  parallelization of the algorithm is one of the directions for further studies. Note that the main calculations in the considered algorithm are performed in two stages: 
data preparation and integration of the ODE system. At the same time, significant memory consumption at the data preparation stage is the main bottleneck limiting a further 
increase of the size of the   model system. In this regard, the parallelization should help satisfy the memory requirements for systems of 
larger sizes (we estimate it as $N \backsimeq 2000$), by using the resources of several nodes of the supercomputer. 
We do not expect a drastic reduction of the computation time; yet it is not the key limiting factor in this case.

Finally, even a mere summation in Eq.~ (7), with a 'dense', randomly generated, rate matrix $A$,  is a heavy computational task already for $N=100$ -- when 
performed on a single node, without accounting for a sparse structure of the matrices. By taking explicitly into account the sparsity and implementing trivial parallelization, 
it was possible to sample over a large ensemble (with more than $10^3$ realizations) of random Lindbladian generators for $N = 100$ and thus explore universal spectral features of the ensemble \cite{Karol}.

	\section{Acknowledgments}\label{acknowledgment}
	
	The authors acknowledge support of the 
	Russian Foundation for Basic Research No. 18-37-00277 (Section \ref{sec:6}),
 President of Russian Federation grant No. MD-6653.2018.2 and Ministry of Education and Science of the Russian Federation Research Assignment No. 1.5586.2017/BY.


\begin{thebibliography}{1}
\bibitem{note} Note them when all dissipative operator are Hermitian, $L^\dagger_p = L_p$, the asymptotic density matrix is trivial normalized identity
	         $\varrho^{A} = \frac{\id}{N}$ (``infinite temperature state``).
\bibitem{lattice} Lattice quantum systems are many-body systems of lattice topologies, i.e., with next-neighbor interaction
	        between its particles or elements.
\bibitem{luch}  An interesting discussion on density matrix parametrization  for spin systems, which accounts also for positivty, can be found in Refs.~\cite{square1,square2}.
	         However, this parametrization is not discussed in the context of evolution, neither unitary nor disispative, so it is no clear how to propagate system in the parameter space.
\bibitem{alternatives} There are at least two different  definitions of the basis based on different normalization conditions. 
	         Here we used the one from Refs.~\cite{alicki0,alicki} (see Ref.~\cite{kimura} for the alternative definition).
\end{thebibliography}
\end{document}